\documentclass[sn-mathphys,onecolumn]{sn-jnl}

\newcommand{\qT}{q_T}
\newcommand{\vqb}{\bar\vq}

\newcommand{\qp}{{\qe }}
\newcommand{\qn}{{q_0}}

 \newcommand{\myskip}[1]{}  

\newcommand{\alb}{\bar\alpha}
\newcommand{\alc}{\alpha_c}

 \newcommand{\vG}{{\bf G}}
 
 \newcommand{\hc}{\hat c}
\newcommand{\hm}{\hat m}

\newcommand{\qb}{\bar q}
\newcommand{\Qb}{\bar Q}

\newcommand{\qe}{q_d}
\newcommand{\uc}{\underline{c}
}\newcommand{\um}{\underline{m}}
 
 \newcommand{\ab}{{\alp\bet}}

\newcommand{\sg}{{\rm  SG}}

\newcommand{\Ie}{I_1}
\newcommand{\It}{I_2}

\newcommand{\Je}{J_1}
\newcommand{\Jt}{J_2}
\newcommand{\Jesq}{J_1^2}
\newcommand{\Jtsq}{J_2^2}

\newcommand{\vq}{{\bf q}}
\newcommand{\vQ}{{\bf Q}}

\newcommand{\bbet}{\bar\beta}

\newcommand{\alp}{\alpha}
\newcommand{\bet}{\beta}
\newcommand{\gam}{\gamma}
\newcommand{\Gam}{\Gamma}

\newcommand{\Lam}{\Lambda}
\newcommand{\lam}{\lambda}

\newcommand{\bq}{\bar q}

\newcommand{\sig}{\sigma}

\newcommand{\vE}{{\bf E}}

\newcommand{\qad}{\hspace{1mm}}

\newcommand{\mmin}{{\hspace{0.2mm}-\hspace{0.2mm}}}
\newcommand{\pplus}{{\hspace{0.2mm}+\hspace{0.2mm}}}

\newcommand{\iss}{{\hspace{0.2mm}=\hspace{0.2mm}}}

\newcommand{\refl}{\ref}

\renewcommand{\d}{{\rm d}}

\newcommand{\half}{{\frac{1}{2}}}

\renewcommand{\max}{{\rm max}}

\newcommand{\vnul}{{\bf 0}}

\newcommand{\tr}{{\rm tr}\,}

\newcommand{\veen}{{\bf 1}}  

\renewcommand{\p}{\partial}   

\newcommand{\BEQ}{\begin{eqnarray}}   
\newcommand{\EEQ}{\end{eqnarray}}   
\newcommand{\BEA}{\begin{eqnarray}}   
\newcommand{\EEA}{\end{eqnarray}}   
\newcommand{\nn}{\nonumber }   
\renewcommand{\d}{{\rm d}}

\newcommand{\cO}{{\cal O}}

\begin{document}

\title{ Hopfield model for patterns with internal structure}

\author{Theodorus Maria Nieuwenhuizen}

\affil*{\orgdiv{Institute for Theoretical Physics}, \orgname{University of Amsterdam}, 
\orgaddress{\street{Science Park 904}, \city{Amsterdam}, \postcode{1090 GL},  
\country{The Netherlands}}}

\abstract{
The spherical version of the Hopfield model for pattern recognition is considered in the static limit. 
Structures inside the patterns are modeled by Gaussian random variables
that reward correlation between pairs of spins in a given pattern.
The free energy is derived analytically with the replica method.  The overlap distribution obeys a self-consistent equation. 
Coming from high temperatures, a spin glass phase is entered, in which patterns and correlations appear at lower temperatures.
For small enough loading capacity, also a glass phase with patterns and correlations appears.
}

\maketitle 

\tableofcontents

\newpage

\section{Introduction}

\hfill{Work dedicated to Desir\'e Boll\'e }

\vspace{3mm}

Analytic descriptions of biological neural networks were introduced for mimicking and understanding the functioning of the human brain.
Their implementation as learning models led to artificial intelligence (AI), which, by machine learning, now has many applications. 
While these schemes employ various ingredients; here we focus on artificial neural networks. 
Often they embody layered neural networks, repetitions  of single neural networks.

Originally, the aim of neural networks was to mimic and understand the functioning of the human brain;
{\color{black} a physical Hebbian update rule was proposed \cite{heerema1999derivation}}.
An important contribution was made by John Hopfield,
who introduced what is now called ``Hopfield models'' or ``Hopfield networks" \cite{hopfield1982neural,bialek2024moving}, 
{\color{black} in which the couplings are taken symmetric.}
In the mean field limit of the model, a system of $N$ artificial neurons, a number $p=\alp N$ of patterns can be  stored, 
where $\alp$ is the loading capacity of the network \cite{amit1987statistical}.

While these are dynamical models,  taking the couplings symmetric assures a statics, 
so that the phase diagram provides insights in the functioning of the model even before addressing the dynamics.
There appears a deep connection with the equilibrium statistical mechanics of spin glasses.
The Sherrington-Kirkpatrick mean field spin glass model \cite{sherrington1975solvable} (SK) was solved 
by Parisi \cite{parisi1980order,mezard1987spin} by introducing infinite order replica symmetry breaking (RSB).
The validity of Parisi's approach was proven by Talagrand\footnote{Giorgio Parisi   was awarded the 2021 Nobel Prize in Physics, 
John Hopfield the one of 2024 and Michel Talagrand the 2024 Abel Prize.}  \cite{talagrand2006parisi}.

The formulation of the SK model for spherical spins provided the first case where the inverse Parisi function $x(q)$ is solvable exactly \cite{nieuwenhuizen1995exactly}.
Spherical spins can be quantized \cite{nieuwenhuizen1995quantum}, so that the model is even well behaved at low temperatures.
Various more complicated functions $x(q)$ occur in site-disordered magnets \cite{nieuwenhuizen1999theory}.
The idea for realizing full RSB in spherical spin models was further implemented and developed,  e.g., in 
\cite{crisanti2004spherical,crisanti2006spherical,crisanti2007equilibrium}.
Spin glass phases with a finite number of replica symmetry breakings, 
{\color{black} sought for but not appearing in site-disordered magnets \cite{nieuwenhuizen1999theory}, 
occur in spherical $s+p$ multispin interaction models \cite {crisanti2007amorphous}.
 }

With the SK couplings turned into a Hopfield couplings, the model was solved by Amit, Gutfreund and Sompolinsky \cite{amit1987statistical}.
It contains a ``glass'' phase with replica symmetry, and at low temperature a spin glass phase with full breaking of replica symmetry.
Well within the glass phase, metastable patterns occur, which at a lower temperature become the lowest free energy states.
This exhibits a retrievable memory.  
 For loading capacity $\alp\le\alp_c\simeq 0.14$ the system provides
an effective retrieval of memory.

There exist many generalizations and applications of SK and Hopfield models; we only mention a few.
Boltzmann machines are SK-type models with external fields; they can be treated as generalized Hopfield networks \cite {marullo2020boltzmann}.  
Restricted Boltzmann machines, having no intralayer couplings, can be employed for many layers and are widely used in machine learning \cite{hinton2006reducing,pozas2021efficient}.
From an other angle, in such models  the effects of social balance on social fragmentation were considered \cite{minh2020effect}.

\subsection{Correlation within patterns}

How the spherical limit of the Hopfield model should be defined, was solved by Boll\'e et al. \cite{bolle2003spherical}. 
It amounts to include a quartic term for the patterns, but of the ``wrong'' sign, as we recall below.

The spherical Hopfield model exhibits a similar phase diagram as the Ising case,  but it has  no full RSB phase.
The theoretical question of how replica symmetry breaking can be incorporated led us to consider patterns with internal correlation. 
This concept can have a broad role beyond the spherical limit and beyond the mean-field limit of the Hopfield model.

Hopfield models with correlated patterns, were discussed, see, e. g., \cite{virasoro1989categorization,fontanari1990storage,tamarit1991pair,lowe1998storage}.
In such models, the parameters $\xi_i^\mu$ of pattern $\mu$ are correlated {\it between different patterns} $\mu'\neq\mu$. 
An example is the remarkable modern habit of making selfies, different photographs of the same person; these are evidently correlated.
In our case, however, the patterns remain $\xi_i^\mu$ uncorrelated, but they are accompanied by a pair correlation $\xi_{ij}^\mu$ {\it within the pattern} $\mu$.
One may think of analyzing weaves in clothes, waves on water surfaces, tracks of birds, city plans, etc.
Clearly, those patterns have  internal structure.
A theory that incorporates this nature may speed up any specific network model.

\subsection{Setup}

The setup of this article is as follows. In section 2 we recall the spherical Hopfield model and make some remarks. 
In section 3 we introduce the correlation in patterns and work out the replica free energy. In section 4 we consider aspects of the phase diagram.
We close with a summary. In  Appendix A we summarize properties of replica symmetry breaking for spherical spin models,
and in Appendix B we analyse the stability of a replica symmetric state with magnetization and correlation.

\renewcommand{\thesection}{\arabic{section}.}
\section{The spherical Hopfield model}
\setcounter{equation}{0}
\renewcommand{\theequation}{\thesection.\arabic{equation}}
\renewcommand{\thesection}{\arabic{section}}

\subsection{The Hamiltonian}

\newcommand{\Jijkl}{J_{ij}^{kl}}
\newcommand{\Jijklmn}{J_{ijkl}^{mn}}

Let us consider the $2+4$ spherical model with  Hamiltonian
\BEQ \hspace{-3mm}
\label{HBolle}
H_B^{(2)}+H_B^{(4)}=-\frac{u_2}{2} \sum_{i,j=1}^N J_{ij}\sig_i\sig_j  \mmin \frac{u_4}{4} \sum_{i,j,k,l=1}^N \Jijkl \sig_i \sig_j \sig_k \sig_l
\EEQ
where the spherical spins are real valued, only constrained by $\sum_{i=1}^N \sig_i^2=N$. 
We add a sixth order interaction,
\BEQ
H_B=H_B^{(2)}+H_B^{(4)}+H_B^{(6)},\quad 
H_B^{(6)}=\frac{u_6}{6} \sum_{ijkmnl} \Jijklmn \sig_i \sig_j \sig_k \sig_l \sig_m \sig_n ,
\EEQ
The case with random, independently distributed Gaussian couplings $J_{\cdots}^{\cdots}$ was shown to have a spin glass phase; it was the first model that allows an exact solution of the 
(inverse) Parisi function $x(q)$ \cite {nieuwenhuizen1995exactly}. 

Here we follow Boll\'e et al. \cite{bolle2003spherical}, to be called  ``ref. B",  and employ the model as a Hopfield model for pattern recognition.
We thus assume $p\equiv\alp N$ patterns of the spins $\sig_i$ ($i=1,\cdots,N$) that are represented, for each pattern $\mu$,
 by Gaussian random variables   $\xi_i^\mu\equiv(${\boldmath{$\xi$}}$^\mu)_i$ of zero average and unit variance.  The couplings read in terms of them
\BEQ  \hspace{-5mm}
J_{ij} \iss \frac{1}{N}\sum_{\mu=1}^p \xi_i^\mu \xi_j^\mu  ,  \qad 
\Jijkl \iss \frac{1}{N^3}\sum_{\mu=1}^p \xi_i^\mu \xi_j^\mu \xi_k^\mu \xi_l^\mu  , \qad
\Jijklmn \iss \frac{1}{N^5}\sum_{\mu=1}^p  \xi_i^\mu \xi_j^\mu \xi_k^\mu \xi_l^\mu \xi_m^\mu\xi_n^\mu.
\EEQ

\subsection{The replica free energy}

In ref. B, the steps to proceed  for the replicated partition sum $\langle Z^n\rangle$ and its logarithm $-\bet N f_n+\cO(N^0)$ have been presented in full detail.
Here we repeat the derivation, so that we can incorporate correlations between the patterns  in next chapter.
In doing so, we suppress the writing of the integration measures,  and only present the successive expressions for the integrand, which is an exponential.
For simplicity of notation, we denote it in the successive steps as $\exp[-\beta Nf_n+\cO(N^0)]$ and present the running expression for $ f_n$.
We can leave out constant terms since they have no physical meaning.

The starting point is to express the Hamiltonian in terms of the order parameter
\BEQ
\um_\mu =    \frac{1}{N}\sum_i      \xi_i  ^\mu \sig_i    ,  
\EEQ
In the partition sum $Z$, factors $1=\int\d m_\mu \delta(\um_\mu-m_\mu)$ are inserted and the $\delta$'s are
expressed  as  $(\bet N/2\pi i)\int_{-i\infty}^{i\infty} \d\hm_\mu \exp[\bet N \hm_\mu(\um_\mu-m_\mu)] $, plane waves over associated fields $\hat m_\mu$;
 they take imaginary values, but their saddle point values will appear to be real. One arrives at
\BEQ
f= \sum_{\mu }[-\frac{u_2}{2} m_\mu^2-\frac{ u_4}{4} m_\mu^4+\frac{u_6}{4} m_\mu^6 +\hat m_\mu ( m_\mu -\frac{1}{N}\sum_i \xi_i^\mu  \sig_i) ] .
\EEQ
Replication leads to summing this over $\alp=1,\cdots,n$, so that $\sig_i\to\sig_i^\alp$,
$m_\mu\to m_\mu^\alp$ and $\hat m_\mu\to \hm_\mu^\alp$.
The quenched $\xi_i^\mu$, identical for each replica, are independent Gaussian random variables with average 0 and variance 1. 
They can be integrated out, leading to
\BEQ
 f_n= \sum_{\mu \alp}[-\frac{u_2}{2}m_\mu^\alp{}^2-\frac{u_4}{4}m_\mu^\alp{}^4 + \frac{u_6}{6}m_\mu^\alp{}^6 + \hat m_\mu^\alp  m_\mu^\alp]  
-\frac{\bet}{2}\sum_{\alp\bet \mu}   \hat m_\mu^\alp \hm_\mu^\bet   {\underline{{\it q}}}_\ab,
\EEQ
  with the overlaps
  \BEQ
   {\underline{q}}_\ab =\frac{1}{N}\sum_i \sig_i^\alp \sig_i^\bet. 
  \EEQ
 For each $\bet=\alp$  this  expresses the spherical constraint $(1/N)\sum_i\sig_i^\alp{}^2=1$.
  Since the $ {\underline{q}}_{\alp\neq\bet}$ can act as order parameters, and also because our later incorporation of correlations produces powers of them,
  we do not yet perform the Gaussian spin integrals, but introduce  similar integrals over $\delta$ functions for the overlaps, 
 again expressed as plane waves integrals over conjugate variables. This leads to
 \BEQ 
 f_n & = &  \sum_{\mu \alp}[-\frac{u_2}{2}m_\mu^\alp{}^2-\frac{u_4}{4}m_\mu^\alp{}^4 + \frac{u_6}{6}m_\mu^\alp{}^6 + \hat m_\mu^\alp  m_\mu^\alp]  
-\frac{\bet}{2}\sum_{\alp\bet \mu}   \hat m_\mu^\alp \hm_\mu^\bet  q _\ab 
\nn\\&& 
-\half \sum_\ab \hat q_\ab q_\ab +\half  \sum_\ab \hat q_\ab \frac{1}{N}\sum_i \sig_i^\alp \sig_i^\bet.
\EEQ
At this moment the spins can be integrated out, and subsequently  the $\hat q_\ab$ by the saddle point method, which replaces the second line successively by
 \BEQ &&
\mapsto -\half \sum_\ab \hat q_\ab q_\ab+
\frac{T}{2} \log \det \hat q \mapsto 
-\frac{T}{2}\log \det  q, 
\EEQ
where we again skip constants.  In the following step, the Gaussian $\hm_\mu^\alp$ with 
$\mu=1,\cdots, p$, where $p=\alp N$, are integrated out.  This results in
 \BEQ
  f_n \iss  \sum_{\mu \alp} \! \big [ \mmin \frac{u_2}{2}m_\mu^\alp{}^2 \mmin \frac{u_4}{4}m_\mu^\alp{}^4 \pplus \frac{u_6}{6}m_\mu^\alp{}^6 \big ]
\pplus \frac{T}{2} \sum_{\mu\ab} m_\mu^\alp(q^{-1})_\ab  m_\mu^\bet
\pplus \frac{\alp \mmin 1}{2} T\tr \log q. \nn \\
\EEQ

Next we assume that pattern $\mu=1$ possibly has a macroscopic average, so we set $m_1^\alp\to m_\alp+m_1^\alp$. The cross terms $m_\alp m_1^\alp$
will vanish at the saddle point value of $m_\alp$, so we can integrate over the $np$  microscopic ``noise'' parameters $m_\mu^\alp\sim 1/\sqrt{N}$. 
(The same result appears when  keeping the integral over $m_1^\alp$.)
This results in the final expression for the replicated free energy per spin,
 \BEQ  \label{bfBollefin}
 &&  f_n= -\frac{u_2}{2}  \sum_{\alp} m_\alp^2- \frac{u_4}{4}  \sum_{\alp} m_\alp^4+ \frac{u_6}{6}  \sum_{\alp} m_\alp^6
  + \frac{T}{2}\sum_{\ab} m_\alp(q^{-1})_\ab  m_\bet
\nn \\ &&
-\frac{T}{2}\log\det q + \frac{\alp T}{2}\log\det(1-\bet_1  q) , \qquad \bet_1=\bet u_2 .
\EEQ
In the following, we take units with $u_2=1$, while at times restoring $u_2$.  

\newcommand{\bT}{\bar T}

\subsection{On the sign of the quartic term}

The reason for choosing the ``wrong'' sign of the quartic term,  $-u_4m^4/4$ in (\ref{HBolle}), becomes clear now.
The cause is that the $m$ and $\qe $ terms stem from the  same physical mechanism.
In the normal case $u_4<0$ there can be a continuous transition with a small 
$m^2\approx  (1 \mmin \qe  \mmin T)/\vert u_4\vert (1 \mmin \qe )$, but near the transition, $\qe \approx 1-T$, the $\alp$ term in (\ref{fqmglass}) has a linear singularity and the argument
of the logarithm becomes negative. This problematic behavior does not occur for $u_4>0$, the case considered in B and  here.
It does come at a cost, as we discuss now.

\subsection{High temperature phase}
\label{highTnom4}

The ``wrong'' sign leads to peculiarities at large $m$.
As seen in B  and in next subsection, one will have $\qe =0$ for $T>1+\sqrt{\alp}$.
 In that case, $ f$  has  extrema  at 
\BEQ \label{mpm=}
m_\pm ^2=\frac{u_4 \pm \sqrt{u_4^2-4u_6(T-1)} }{2u_6 },
\EEQ
which are real and possibly problematic for $T<1+u_4^2/4u_6$. 
The ones at $m=\pm m_-$ are unstable and can be discarded; the ones at $m=\pm m_+$ are stable and need consideration. 
For small $u_6$ one has a large $m_+ $, which is a peculiarity of the model. Indeed, for Ising spins $\sig_i=\pm1$ one has the upper bound
\BEQ \label{mmax}
\um_\mu  =\frac{1}{N}  \sum_{i=1}^N \xi_i^\mu \sig_i\ \le \vert \um_\mu\vert \le  m_\max=\frac{1}{N}  \sum_{i=1}^N  \langle  \vert \xi_i ^\mu\vert \rangle = \sqrt{\frac{2}{\pi}}   
\quad\text{(Ising)}. 
\EEQ
For spherical spins, the maximum is larger. With help of a Lagrange multiplier $\lam(N-\sum_i\sig_i^2)$ one finds 
$\sig_i^\max= \xi_i^\mu (N/\sum_j\xi_j^\mu{}^2)^{1/2}$ so that  $m_\max=1$.

States $m=\pm\, m_+$ for $m_+>1$  are artifacts,
hence we neglect them from now on. Keeping this in mind, we set $u_6=0$, as was done in B.
Another option would be to restrict the model to $0<u_6<u_4^2/4\sqrt{\alp}$, which makes the $m_\pm$ roots (\ref{mpm=}) 
complex in this domain and hence physically irrelevant.

\subsection{Free energy in the glass phase}

The replica symmetric phase is called ``glass phase''.  At some low temperature it may host retrievable patterns. 
It  involves a common order parameter $\qe $ for all $q_{\alp\neq\bet}$,  
\BEQ \label{glass}
q_{\alp\bet}=(1-\qe )\delta_{\alp\bet}+\qe  E_{\alp\bet}, \quad (E_{\alp\bet}=1,\quad \alp,\bet =1,\cdots,n).
\EEQ
The eigenvalues, eigenvectors and their degeneracies of this $n\times n$ matrix are
\BEQ
&& \lam_1=1+(n-1)\qe , \qquad e^{(1)}_\alp=\frac{1}{\sqrt{n}} ,  \qquad {\rm deg}=1 ,  \nn \\ 
&& \lam_j=1-\qe  ,\qquad e^{(j)}_\alp=\frac{1-n\delta_{\alp,j}}{\sqrt{n(n-1)}}  , \qquad {\rm deg}=n-1.
\EEQ

The patterns will be replica symmetric, $m_\alp=m$, so there results
\BEQ  \label{fqmglass}
&&
 f_n=-\frac{u_2}{2} n m ^2 - \frac{u_4}{4}n m^4
+\frac{nTm^2}{2[1+(n-1)\qe ]}  \nn\\&& 
-\frac{T}{2}\Big\{ \log[1+(n-1)\qe ]+(n-1)\log(1-\qe ) \Big\} \nn\\&&
+\frac{\alp T}{2} \Big\{  \log\frac{T-[1+(n-1)\qe ]}{T}+ (n-1)\log\frac{T-(1-\qe )}{T} \Big\} .
\EEQ           
This is valid for all integer $n$. In the replica method, one assumes that the limit $n\to0$ can be taken by analytic continuation.
The free energy per replicated spin,  $f=f_n/n$, thus becomes
\BEQ  \label{fqmglass}
f &=&   -\frac{u_2}{2}m^2 - \frac{u_4}{4}m^4  +\frac{T m^2}{2(1-\qe )} 
-\frac{T}{2}\Big\{\log(1-\qe )+\frac{\qe }{1-\qe }\Big\}
\nn \\&&
+\frac{\alp T}{2} \Big\{ \log[1-\bet(1-\qe )]-\frac{\qe }{T-1+\qe } \Big\}. 
\EEQ           
We take units where  $u_2=1$.
The mean field equations are
\BEQ \label{mfBolle} \hspace{-4mm}
\frac{T \mmin 1\pplus \qe  }{1-\qe } m=u_4m^3 
, \quad 
\frac{\qe }{(1-\qe )^2} = \frac{m^2}{(1-\qe )^2} +\frac{\alp \qe }{(T \mmin 1 \pplus \qe )^2} .
\EEQ
At  $m=0$, the solution of (\ref{mfBolle}) reads for $T<T_g$
\BEQ
\qe =1-\frac{T}{T_g},\qquad T_g=1+\sqrt{\alp}.
\EEQ 
This is the pure glass phase. 
A glass phase with pattern recovery occurs when
\BEQ
m=\sqrt{\frac{T-1+\qe }{u_4(1-\qe )}} .
\EEQ
Inserting this in (\ref{mfBolle}) leads to a quartic equation for $\qe $.
The full phase diagram was analyzed in ref. B.
Well inside the glass phase, patterns become metastable and, at still lower temperature, they become stable.

\renewcommand{\thesection}{\arabic{section}.}
\section{Incorporating structure inside patterns}
\setcounter{equation}{0}
\renewcommand{\theequation}{\thesection.\arabic{equation}}
\renewcommand{\thesection}{\arabic{section}}

We extend the spherical neural net model with correlations due to structure inside the patterns.
The same idea can be applied to many other models, including those with Ising spins.

\subsection{The correlation Hamiltonian}

As a further step to generalize the model, we assume a Hamiltonian for correlations inside $p_c$ of the patterns,
with $p_c\le p$ and $\alc=p_c/N\le\alp$, 
\BEQ \hspace{-5mm} 
H_c^{(2)} \iss \mmin \frac{v_2}{2N^2}\sum_{\mu=1}^{p_c} \sum_{ijkl=1}^N  \xi_{ij}^\mu \xi_{kl}^\mu \sig_i \sig_j \sig_k \sig_l 
\iss \mmin \frac{v_2}{2} N \sum_{\mu} \uc_\mu^2, \quad   
\EEQ
  for independent random Gaussian variables $\xi_{ij}^\mu$ for all $i,j=1,\cdots , N$ with average 0 and variance 1.
  In the standard situation $p_c=p$, so that $\alc=\alp$, each pattern can have structure in principle.
 However,  for instructive reasons we will consider general $p_c$ and $\alc$.
The pattern correlation is introduced as
\BEQ
\uc_\mu  \iss \frac{1}{N^{3/2}}\sum_{i,j=1}^N \xi_{ij}^\mu \sig_i \sig_j  
  =\frac{1}{N^{3/2}}\sum_{i<j} (\xi_{ij}^\mu+\xi_{ji}^\mu) \sig_i \sig_j +\frac{1}{N^{3/2}}\sum_{i} \xi_{ii}^\mu \sig_i ^2,
\EEQ
The combination $\xi_{ij}^\mu+\xi_{ji}^\mu$ is a Gaussian with average 0 and variance $\sqrt{2}$. 

For Ising spins one has, similar to (\ref{mmax}), the upper limit,
\BEQ \label{cmax}
\vert \uc_\mu \vert = \frac{1}{N^{3/2}} \big\vert  \sum_{ij}\xi_{ij}^\mu \sig_i \sig_j  \big \vert\le 
\frac{1}{N^{3/2}} \sum_{ij} \vert \xi_{ij}^\mu  \vert =  \sqrt{  \frac{2N}{\pi}}  .
\EEQ
This makes the existence  plausible of condensed patterns with $c_\mu=\cO(1)$, even for spherical spins;
like the non-condensed $\um_\mu$ of previous section, the non-condensed $\uc_\mu\sim N^{-1/2}$ are noise effects, to be averaged over (integrated out).

To model condensation of pattern correlations, we consider, analogous to subsection \ref{highTnom4}, the higher order terms
\BEQ \hspace{-3mm} && 
H_c^{(4)}  = -\frac{v_4}{4N^5}\sum_{\mu} \Big(\sum_{ij}  \xi_{ij}^\mu \sig_i \sig_j \Big)^4 = -\frac{v_4}{4}N\sum_\mu \uc_\mu^4 , \quad 
H_c^{(6)}  =\frac{v_6}{6}N\sum_\mu \uc_\mu^6 , 
\nn\\&& 
H_{cm}^{(4)}=
-\frac{w_4}{2N^4} \Big(\sum_{i}  \xi_{i}^\mu \sig_i \Big)^2 \Big(\sum_{ij}  \xi_{ij}^\mu \sig_i \sig_j \Big)^2
=-\frac{w_4}{2} N\sum_\mu \um_\mu^2 \,  \uc_\mu^2  , \qquad 
 \EEQ
(Odd powers of $\uc_\mu$ are odd in $\xi_{ij}^\mu$ and omitted.)
Taking $v_{2,4,6}>0$, the $H_c^{(2,4)}$ reward correlations, while $H_c^{(6)}$ prevents too large ones.
For $w_4>0$ the term $H_{cm}^{(4)}$ rewards correlation $\uc_\mu$ in the  pattern $\um_\mu$.
For $w_4=0$, stable high temperature minima $\pm c_+$ occur with $c_+ > v_4/2v_6$. For $v_6<v_4\sqrt{\pi/8 N}$ they exceed the Ising bound (\ref{cmax}).
We neglect them  and also set $v_6\to0$.
We  thus collect
\BEQ
\frac{H_c}{N}=\sum_{\mu=1}^{p_c}\Big(-\frac{v_2}{2}\uc_\mu^2-\frac{v_4}{4}\uc_\mu^4
-\frac{w_4}{2}\um_\mu^2\,\uc_\mu^2\Big) .
\EEQ

  \subsection{Contribution to the replicated free energy}
  
In the replicated partition sum, we follow the setup of section 2 and introduce the additional order parameters $c_\mu^\alp$ and the associated  $\hc_\mu^\alp$,
to get the correlation contribution to the free energy 
\BEQ \hspace{-3mm}
 f_n^c= \sum_{\mu\alp}\Big [ - \frac{v_2}{2}c_\mu^\alp{}^2  -\frac{v_4}{4} c_\mu^\alp{}^4-\frac{w_4}{2}\um_\mu^\alp{}^2 c_\mu^\alp{}^2
 \pplus   \hc _\mu^\alp (c_\mu^\alp  \mmin \frac{1}{N^{3/2}}\sum _{ij}\xi^\mu_{ij}\sig_i^\alp \sig_j^\alp) \Big] .
\EEQ
We integrate out the quenched $\xi_{ij}^\mu$. In the result we may replace $(1/N)\sum_i\sig_i^\alp\sig_i^\bet$ by $ q_\ab$ 
because of the $\delta$ functions already present in the part treated in section 2; 
for the same reason, $\um_\mu^\alp$ can be replaced by $m_\mu^\alp$.
 This leads to
\BEQ
 f_n^c= \sum_{\mu\alp}\Big [ - \frac{v_2}{2}c_\mu^\alp{}^2  -\frac{v_4}{4} c_\mu^\alp{}^4
 -\frac{w_4}{2} m_\mu^\alp{}^2 c_\mu^\alp{}^2 \pplus   \hc _\mu^\alp c_\mu^\alp  \Big]
 -\frac{\bet}{2}  \sum_{\mu\alp\bet} \hc_\mu^\alp \hc_\mu^\bet Q_{\alp\bet} ,
 \EEQ
 where 
 \BEQ
 Q_{\alp\bet}=  (q_{\alp\bet})^2 , \qquad (\alp,\bet=1,\cdots,n),
 \EEQ 
is evidently different from $(q^2)_{\alp\bet}=q_{\alp\gam}q_{\gam\bet}$; it takes the place of $q_\ab$ in section 2.
 The next step is to integrate out the $\hc$, which leads to
\BEQ \hspace{-3mm} &&
 f_n^c= \sum_{\mu\alp}\Big [ - \frac{v_2}{2}c_\mu^\alp{}^2  -\frac{v_4}{4} c_\mu^\alp{}^4-\frac{w_4}{2} m_\mu^\alp{}^2 c_\mu^\alp{}^2
  \Big]
 \pplus \frac{T}{2}  \sum_{\mu\alp\bet} c_\mu^\alp c_\mu^\bet (Q^{-1})_{\alp\bet} \nn\\&&
\pplus \frac{\alc T}{2} \tr \log  Q .
 \EEQ
Finally,  assume that the $\mu\iss1$ pattern $m_1^\alp\equiv m_\alp$ and its correlation $c^1_\alp\equiv c_\alp$ are possibly of order unity, 
and integrate out the other $c_\mu^\alp$.
This leads to 
\BEQ \label{bfcfin}
\hspace{-3mm}
  f_n^c &=&   \sum_{\alc} \Big[-\frac{ v_2}{2} c_\alp^2  \mmin \frac{v_4}{4} c_\alp^4  \mmin \frac{w_4}{2} m_\alp^2c_\alp^2 \Big] 
 +  \frac{T}{2}\sum_{\alp\bet} c_\alp (Q^{ \mmin 1})_{\alp\bet} c_\bet  \nn\\&&   + \frac{\alc T }{2}  \log\det(\veen \mmin \bet_2 Q) ;
 \qquad (\bet_2\equiv \bet v_2=\frac{v_2}{T}) .
 \EEQ

\subsection{The full free replicated energy}

Combining eqs. (\ref{bfBollefin}) and (\ref{bfcfin}),  the total  replicated free energy $f_n$ reads
\BEQ  \label{fnfin}
 f_n &=&  \sum_\alp \Big[
 - \frac{u_2}{2} m_\alp^2-\frac{ u_4}{4} m_\alp^4
 -\frac{v_2}{2}  c_\alp^2  - \frac{ v_4}{4}  c_\alp^4  -\frac{w_4}{2}m_\alp^2c_\alp^2\Big] \nn \\ && 
 +\frac{T}{2} \sum_{\alp\bet} \Big [ m_\alp   (q^{-1})_{\alp\bet}  m_\bet  +  c_\alp  (Q^{-1})_{\alp\bet} c_\bet\Big] 
\\&&  +\frac{\alp T}{2} \log \det (\veen-\beta_1 q) + \frac{\alc T}{2} \log \det (\veen-\bet_2 Q)  - \frac{T}{2} \log \det q . \nn
\EEQ

Objects with one replica index must be replica symmetric, viz. $m_\alp=m$ and $c_\alp=c$,
while $q_{\alp\bet} $ and $Q_{\alp\bet} $ can exhibit replica symmetry breaking.
So the free energy per spin $f=F/N$ becomes the limit $n\to 0$ limit of
\BEQ \label{bfgen}
f &=& - \frac{u_2}{2} m^2-\frac{ u_4}{4} m^4  -\frac{v_2}{2}  c^2  - \frac{ v_4}{4}  c^4  -\frac{w_4}{2}m^2c^2 \nn \\ && 
 +\frac{T}{2n} \sum_{\alp\bet} \Big [ m^2   (q^{-1})_{\alp\bet}   +  c^2  (Q^{-1})_{\alp\bet} \Big] 
\\&&  +\frac{\alp T}{2n} \log \det (\veen-\beta _1 q) + \frac{\alc T}{2n} \log \det (\veen-\bet_2 Q)  - \frac{T}{2n} \log \det q . \nn
\EEQ
where, restoring general $u_2$,
\BEQ
\beta_1=\bet u_2, \quad T_1=\frac{1}{\beta_1}=\frac{T}{u_2},\qquad 
\beta_2=\beta v_2,\qquad T_2=\frac{1}{\beta_2}=\frac{T}{v_2}.
\EEQ

\subsection{The free energy in the replica limit $n\to0$}

For general replica symmetry breaking, the terms of (\ref{bfgen}) are analyzed in the Appendix, and the limit $n\to 0$ is taken. 
Collecting the results leads to
\BEQ \label{fphysRSB}
 f = && - \frac{u_2}{2}m^2-\frac{u_4}{4}m^4 -\frac{v_2}{2}c^2-\frac{v_4}{4}c^4 -\frac{w_4}{2}m^2c^2 +\frac{T m^2}{2\Ie(q_0)}+\frac{T c^2}{2I_2(q_0)}
\nn\\&& 
+\frac{\alp T}{2}\Big\{  \log[1-\bet(1-\qe )]- \frac{q_0}{\Je(q_0)}  - \int_{q_0}^{\qe }\frac{\d q}{\Je(q)  }  \Big \} \nn \\ &&
+\frac{\alc T}{2}\Big\{   \log[1-\bet_2(1-\qe ^2)] -\frac{q_0^2}{\Jt(q_0)}-  \int_{q_0}^{\qe }\frac{\d q\,2q}{\Jt(q)  }  \Big \}  
\nn\\ &&
-\frac{T}{2} \Big\{  \log(1-\qe )+ \frac{q_0}{\Ie(q_0)} +\int_{q_0}^{\qe }\frac{\d q}{ \Ie(q) }  \Big \}  ,
\EEQ
where
\BEQ
\Je(q)=T_1 -\Ie(q),\qquad \Jt(q)=T_2-I_2(q).
\EEQ
  $\Ie$ and $\It$ involve the Parisi function $x(q)$ with $x(q)=1$ for $\qe \le q\le1$,
\BEQ \label {II2def} 
\hspace{-7mm}
I_1(q)  \iss  \! \int_{q}^1 \d \qb \, x(\qb)\, \iss 1 \mmin  \qe  \pplus \! \int_q^{\qe }\d \qb\, x(\qb),   \quad  
I_2(q) \iss 1 \mmin  \qe^2  \pplus \! \int_q^{\qe }\d \qb\, 2\qb x(\qb) ,  \qad  
\EEQ
It holds that
\BEQ
\Ie'(q)=-x(q) ,\qquad I_2'(q)=-2qx(q) . 
\EEQ

{\color{black}

\subsection{Replica symmetric free energy}

As in ref. B, a replica symmetric ``glass'' phase may occur.  Hereto one sets $q_0\to q_d$, while $x(q) = 0$ for $0\le q\le q_d$ and $x(q)=1$ for $q_d\le q \le 1$, 
so that $\Ie(q)\to 1-\qe $,  $I_2(q)\to 1-\qe ^2$. This simplifies (\ref {fphysRSB}) to
\BEQ  \label{fqmglassz}
&& \nn
f=- \frac{1}{2}m^2-\frac{u_4}{4}m^4 -\frac{v_2}{2}c^2-\frac{v_4}{4}c^4 -\frac{w_4}{2}m^2c^2
+\frac{Tm^2}{2(1-\qe )} +\frac{Tc^2}{2(1-\qe ^2)} 
\\&&
+\frac{\alp T}{2} \Big\{ \log[1-\bet(1-\qe )]-\frac{\qe }{T-1+\qe } \Big\} 
- \frac{T}{2} \Big\{\log(1-\qe )+\frac{\qe }{1-\qe } \Big\} 
\nn \\ && 
+\frac{\alc T}{2} \Big\{ \log[1-\bet_2(1-\qe ^2)]-\frac{\qe ^2}{T_2-1+\qe ^2} \Big\}. 
\EEQ           
In the limit $\alc\to0$ and $c\to0$, the free energy without correlations of ref B,  reproduced in eq. (\ref{fqmglass}), is recovered.

} 
 
\renewcommand{\thesection}{\arabic{section}.}
\section{Mean field equations}
\setcounter{equation}{0}
\renewcommand{\theequation}{\thesection.\arabic{equation}}
\renewcommand{\thesection}{\arabic{section}}

In the high temperature phase one has $\qe =x(q)=0$,  $\Ie(q)=I_2(q)=1$, and $m=c=0$.

\subsection{The pattern and its correlation}
\label{sec4.1}

The mean field equations for $m$ and $c$ are
\BEQ\label{mceq}
\big[ K_1  -w_4c^2  \big]m= u_4m^3,\qquad
\big[ K_2  -w_4m^2  \big]c= v_4 c^3 .
\EEQ
where
\BEQ \hspace{-3mm}
K_1=u_2\frac{\Je(q_0)}{\Ie(q_0)} =\frac{T}{\Ie(q_0)}-u_2,\quad
K_2=v_2\frac{\Jt(q_0)}{I_2(q_0)}=\frac{T}{I_{2}(q_0)}-v_2.
\EEQ
They have the  solution that both $m=c=0$, or one of them being finite
\BEQ \label{mc0m0c}
m=\pm\sqrt{ \frac{K_1}{u_4 }} , \quad c=0 ;   \qquad \text{or} \qquad 
c=\pm\sqrt{ \frac{K_2 }{ v_4 } } , \quad m=0,
\EEQ
that get stabilized when the spin glass is strong enough. At temperatures well below these two,  they can both be finite,
  \BEQ   \label{mc}
  && \hspace{-3mm} 
m^2 \iss \frac{ v_4K_1   \mmin w_4 K_2  }{ u_4v_4-w_4^2 } , \qquad  
c^2 \iss \frac{ u_4 K_2 (q_0)   \mmin w_4 K_1 }{ u_4v_4-w_4^2 } .
 \EEQ
A small $c$ occurs when  $u_4K_2\approx w_4K_1$, whence  $m^2=(K_1-w_4c^2)/u_4$
smoothly joins the first case of (\ref{mc0m0c}).
Likewise,  small $m$  connects to the second case.
These cases apply to $u_2>v_2$ and $u_2<v_2$, respectively.

\subsection{The glass phase}

A replica symmetric ``glass'' phase may exist. 
\myskip{ One sets $\Ie(q)\to 1-\qe $,  $I_2(q)\to 1-\qe ^2$  and $q_0\to \qe $.
This simplifyies (\ref {fphysRSB}) to
\BEQ  \label{fqmglassz}
&& \nn
f=- \frac{1}{2}m^2-\frac{u_4}{4}m^4 -\frac{v_2}{2}c^2-\frac{v_4}{4}c^4 -\frac{w_4}{2}m^2c^2
+\frac{Tm^2}{2(1-\qe )} +\frac{Tc^2}{2(1-\qe ^2)} 
\\&&
+\frac{\alp T}{2} \Big\{ \log[1-\bet(1-\qe )]-\frac{\qe }{T-1+\qe } \Big\} 
- \frac{T}{2} \Big\{\log(1-\qe )+\frac{\qe }{1-\qe } \Big\} 
\nn \\ && 
+\frac{\alc T}{2} \Big\{ \log[1-\bet_2(1-\qe ^2)]-\frac{\qe ^2}{T_2-1+\qe ^2} \Big\}. 
\EEQ           
}
The expressions for $m$ and $c$ are given in sec. \ref{sec4.1}, where now
\BEQ
K_1=\frac{T}{1-q_d} -u_2,\qquad K_2=\frac{T}{1-q_d^2}-v_2.
\EEQ
The mean field equation for $\qe $ reads
\BEQ \label{mfBolleQ}
\frac{\qe }{(1-\qe )^2} = \frac{m^2}{(1-\qe )^2} +\frac{2\qe c^2}{(1-\qe ^2)^2}
+\frac{\alp \qe }{(T-1+\qe )^2}  +\frac{2\alc \qe ^3}{(T_2-1+\qe ^2)^2} .
\EEQ

\subsection{The spin glass phase}

We now consider a nontrivial function $x(q)$ and vary $f$ with respect to $x(q)$,  employing  $\delta \Ie(\qb)/\delta x(q) =\theta(q-\qb)$ 
and $\delta I_2(\qb)/\delta x(q) = 2q\theta(q-\qb)$ from (\ref{II2def}). This  yields for $q_0<q<\qe $
\BEQ  \label{qqRSB}
&&
Z(q)\equiv 
\frac{m^2}{\Ie^2(q_0)} +   \frac{2qc^2}{\It^2(q_0)}  
 + \frac{\alp q_0 }{ \Jesq(q_0)  }  +\frac{2\alc q_0^2q}{\Jtsq(q_0) }  -\frac{q_0}{\Ie^2(q_0)} +
 \nn\\ &&
\alp  \int_{q_0}^{q} \frac{\d \qb  }{  \Jesq(\qb ) } 
+2\alc q \int_{q_0}^{q}\frac{\d \qb  \,2 \qb  }{ \Jtsq (\qb  ) }
  -\int_{q_0}^{q}\frac{\d \qb  }{\Ie^2(\qb  ) }  
 = 0 .
\EEQ
It can be verified that $Z(q_0)=0$ and $Z(\qe)=0$ coincide with  $\p f/\p q_0=0$ and $\p f/\p \qe=0$, respectively.
The first case reads explicitly
\BEQ
Z(q_0)=
\frac{m^2}{\Ie^2(q_0)} +   \frac{2q_0c^2}{\It^2(q_0)}  
 + \frac{\alp q_0 }{ \Jesq(q_0)  }  +\frac{2\alc q_0^3}{\Jtsq(q_0) }  -\frac{q_0}{\Ie^2(q_0)} =0.
\EEQ
Taking the derivative of (\ref {qqRSB}) with respect to $q$ yields
\BEQ \label{Zpq=}
\hspace{-5mm}
Z'(q)= \frac{2c^2}{\It^2(q_0)}   \pplus   \frac{\alp }{  \Jesq(q) }   \pplus \frac{2\alc q_0^2}{\Jtsq(q_0) }   
 \pplus \frac{4\alc q^2 }{ \Jtsq (q  ) }  - \frac{1 }{\Ie^2(q  ) }   \pplus 2\alc  \int_{q_0}^{q}\frac{\d \qb  \,2 \qb  }{ \Jtsq (\qb  ) }
 = 0 .
\EEQ
At $q_0$ this imposes a second boundary condition
\BEQ \label{Zpq0}
\hspace{-5mm}
Z'(q_0)= \frac{2c^2}{\It^2(q_0)}  +  \frac{\alp }{  \Jesq(q_0) } +\frac{6\alc q_0^2}{\Jtsq(q_0) }     - \frac{1 }{\Ie^2(q_0  ) } =0,
\EEQ
which makes $Z(q_0)$ equivalent to
\BEQ \label{Zsq0}
Z(q_0)=
\frac{m^2}{\Ie^2(q_0)}   -\frac{4\alc q_0^3}{\Jtsq(q_0) }  =0.
\EEQ 
In $Z''(q)$ we insert $\Ie'(q)=-x(q)$ and $\It'(q)=-2qx(q)$.  Solving $Z''=0$ yields 
\BEQ \label{xq=}
x(q)=q\times \frac{ 6 \alc \Ie^3(q) \Je^3(q) \Jt(q) } {   \Je^3(q)\Jt^3(q) + \alp \Ie^3(q) \Jt^3(q) + 8 \alc q^3 \Ie^3(q)  \Je^3(q)}  .
\EEQ
With the definitions (\ref{II2def}), this is a self-consistent equation for $x(q)$.
With the conditions (\ref{Zpq0}) and (\ref{Zsq0}), the problem is well posed.
Taking $m$ and $c$ from any of the four cases in section \ref{sec4.1}, eqs.  (\ref{Zpq0}) and  (\ref{Zsq0}) determine the endpoints 
$q_0$ and $\qe$ of $x(q)$.  When  $m=0$, also $q_0=0$, as usual in spin glasses.

\renewcommand{\thesection}{\arabic{section}.}
\section{Aspects of the phase diagram}
\setcounter{equation}{0}
\renewcommand{\theequation}{\thesection.\arabic{equation}}
\renewcommand{\thesection}{\arabic{section}}

\subsection{Spin glass phase from high $T$}

Near a phase transition from high temperature,  the pattern overlap $m$, the pattern correlation $c$ and the replica overlaps $q_{\alp\neq\bet}$ are small or still zero.
Remembering that $q_{\alp\alp}=1$,  we set $q_{\alp\bet}=\delta_{\alp\bet}+\tilde q_{\alp\bet}$ and $Q_{\alp\bet}=\delta_{\alp\bet}+\tilde Q_{\alp\bet}$
with $\tilde q_{\alp\alp}=\tilde Q_{\alp\alp}=0$ and $\tilde Q_\ab=\tilde q_\ab^2$.
We define
\BEQ
\bet=\frac{1}{T},\qad 
\beta_2=\frac{1}{T_2}=\beta v_2, \quad 
\bbet=\frac{\bet}{1 \mmin \bet} =\frac{1}{T \mmin 1},  \quad 
\bbet_2=\frac{\bet_2}{1 \mmin \bet_2} =\frac{v_2}{T \mmin  v_2} .
\EEQ
Assuming that $m=c=0$, expansion of (\ref{bfgen}) up to order $\tilde q ^3$ and $\tilde Q^2$ results in  
\BEQ \label{bfqQser}
f= \frac{T}{2n}\,  \Big[ \frac{1-\alp\bbet^2}{2} \tr \tilde q^2 -\frac{1+\alp\bbet^3}{3} \tr \tilde q^3 -\frac{\alc \bbet_2^2}{2} \tr \tilde Q^2 \Big]  
  , \qad \tr \tilde Q^2=\sum_\ab q_\ab^4  .
\EEQ
This is just the shape of Parisi's free energy functional near the transition at $T_\sg=1+\sqrt{\alp}$, where $\alp \bar\bet^2 \mmin 1\sim T \mmin T_\sg$.
In particular, the quartic replica symmetry breaking term has the negative sign of the SK model.

Hence coming from high $T$, the physical  state has full replica symmetry breaking  due to the presence of the
correlation parameter $\alc$ in (\ref {bfqQser}), that is to say, due to the presence of non-condensed patterns.
With $m=q_0=0$ still and, for $v_2<1+\sqrt{\alp}$ small enough, also $c=0$. Then eq. (\ref{Zpq0}) solves as
\BEQ
\Ie(0) =\frac{T}{1+\sqrt{\alp}} , \qquad \Je(0)=\frac{T\sqrt{\alp}} {1+\sqrt{\alp}},\qquad T_\sg=1+\sqrt{\alp}  .
\EEQ
With $\Ie(q)=1$ at high $T$, this indicates a phase transition to a spin glass phase at $T_\sg$. 
These values imply that $x(q)/q$ jumps at $T_\sg$ to a  finite value.
Indeed, at $T_\sg$ one has from (\ref{xq=})  with $\Ie(0)=I_2(0)=1$,
\BEQ
x(q) =\frac{6\alc \sqrt{\alp} (T_2-1) \, q} { (1+\sqrt{\alp}) (T_2-1)^3 + 8\alc \sqrt{\alp} \, q^3  } ,\qquad T_2=\frac{1+\sqrt{\alp}}{v_2}.
   \EEQ
   This holds for $0\le q\le q_d\sim 1-T/T_\sg\ll 1$, so that  the $q^3$ term can be neglected.
 It follows  that $x(q)/q$ is a constant, finite for $T\to T_\sg$, in this narrow $q$-region, as it happens for the Parisi solution of the SK model.

\subsection{Stability at a glass  to spin glass transition}

In this subsection we consider the possibility of a (marginally) stable glass phase,
a replica symmetric phase with possibly two magnetization-type order parameters (the magnetization and the correlation),  making use 
of results for the random Blume-Emery-Griffiths-Capel model by Crisanti and Leuzzi \cite{crisanti2004thermodynamic}. 

In eq. (\ref{glass}), we add a replica symmetry broken perturbation $\qb_{\alp\bet}=\qb_{\bet\alp}$,
\BEQ
&& \, q_{\alp\bet}  \, =(1-\qe )\delta_{\alp\bet}+\qe  E_{\alp\bet} +\qb_{\alp\bet} , \quad (E_{\alp\bet}=1,\quad \alp,\bet =1,\cdots,n) , 
\nn \\ &&
Q_{\alp\bet}=(1-\qe^2 )\delta_{\alp\bet}+\qe^2  E_{\alp\bet} +\Qb_{\alp\bet} , \quad 
\Qb_\ab=2\qe \qb_\ab+\qb^2_\ab ,
\EEQ
with $\qb_{\alp\alp}=\Qb_{\alp\alp} = 0 $.   

We expand (\ref{fnfin}) to second order in $(\vqb)_\ab=\qb_{\alp\bet} $, while $\bar\vQ=2\qe\vqb+\vqb_2$ with $(\vqb_2)_\ab=\qb_\ab^2$.
 The terms linear in $\vqb$ cancel at the saddle point.
The second order fluctuation matrix can be diagonalized by the method of \cite{morone2014replica}; we restrict ourselves  to 
the dangerous mode,  the ``replicon'' (or ``ergodon'') characterized by $\sum_\bet \bar q_\ab=0$, in matrix form: $\vE\vqb=\vnul$.

A detailed analysis is discussed in Appendix B. 
The quadratic replicon perturbations  around the glass thus combine as $(T/2)\Gam \sum_\ab\qb_\ab^2$ with
 \BEQ \label{Gam=}
\Gam =
\frac{1}{2(1\mmin \qe)^2}
 \mmin \frac{c^2}{(1\mmin \qe^2)^2}
 \mmin \frac{\alp } {2 (T\mmin 1\pplus\qe)^2 } 
\mmin \frac{3\alc \qe^2  }{(T_2\mmin 1\pplus\qe^2)^2 }   .
\EEQ
When $q_d$ is nonzero, we may insert the mean field equation (\refl{mfBolleQ}), leading to  the simpler  form
\BEQ
\Gam = \frac{1}{2\qe}\Big[ \frac{m^2}{(1 - \qe)^2}  -  \frac{4\alc \qe^3 }{(T_2-1 + \qe^2 )^2} \Big].
    \EEQ
The condition $\Gam \downarrow0$ sets  a smooth transition from a glas phase to a spin glass phase.
For $\alc=0$, there is no region with $\Gam<0$, hence no spin glass  \cite{bolle2003spherical}.
Notice that due to eq. (\ref{Zsq0}), this combination vanishes at the border where a spin glass smoothly goes to a glass 
by having $x(q)\to0$ for $q<\qe$.

A second important eigenvalue is the $\Lam_-$ of eqs. (\ref{Lammin}) and (\ref{Lammin0}).

\subsection{The glass phase at $T=0$}

Simplifications occur in the $T\to0$ limit. At small $T$  we set
\BEQ
\qe=1-\frac{T}{y+1}+\cO(T^2) , \qquad \frac{T}{T-1+\qe}=\frac{y+1}{y}+\cO(T)
\EEQ
Eq. (\ref{fqmglassz}) leads to the $T=0$ free energy of the glass
\BEQ \label{f0gl}
f_0&=&\frac{ y}{2}m^2 -\frac{u_4}{4}m^4 -\frac{v_4}{4}c^4 -\frac{w_4 }{2}c^2 m^2
+\frac{1 - 2 v_2 + y}{4}c^2
\nn\\ &&
-\frac{\alp (1 + y)}{2 y}
-\frac{\alc v_2 (1 + y)}{2(1- 2 v_2 +  y)}
-\frac{1 + y}{2}.
\EEQ
The squares of the order parameters of its patterns become
\BEQ
m^2=\frac{ 2  v_4 y -  w_4( y+1-2v_2)}{ 2 (u_4 v_4 - w_4^2 )} ,\quad  
c^2=\frac{ u_4(y+1-2v_2) - 2w_4 y}{ 2 (u_4 v_4 - w_4^2) } ,
  \EEQ
  or either of them vanishing, obtained by setting $w_4\to0$ and discarding the other one, or both zero.
The mean field equation for $y$ is
\BEQ \label{mfg1}
m^2+\half c^2+\frac{\alp}{y^2}+\frac{2\alc v_2^2}{(1-2v_2+y)^2} =1 .
\EEQ
The replicon has eigenvalue
\BEQ \hspace{-6mm} \label{ergodonT0}
\Gam_0=\frac{(1+y)^2}{2}\big[m^2-
\frac{4\alc v_2^2}{(1-2v_2+y)^2}\Big]. 
\quad
\EEQ
At given set of parameters, one can solve (\ref{mfg1}) for $y$ and get $\Gam_0$ from (\ref{ergodonT0}).

To get insight in the problem, we  rewrite the equations, keeping in mind that we focus on small $\Gam_0$.
The situation simplifies for $v_2=\half$, where (\ref{f0gl}) becomes, apart from a constant,
\BEQ \label{f0glv2half} \hspace{-5mm}
f_0=\frac{ y}{2}m^2 -\frac{u_4}{4}m^4 -\frac{v_4}{4}c^4 -\frac{w_4 }{2}c^2 m^2
+\frac{y}{4}c^2
-\frac{1+\alb}{2 y}  , \quad \alb=\alp+\half \alc .
\EEQ
The pattern parameters read
\BEQ \label{m2c2v2ha}
m^2=\frac{ 2  v_4  -  w_4 }{ 2 (u_4 v_4 - w_4^2 )} y , \qquad 
c^2=\frac{ u_4- 2w_4}{ 2 (u_4 v_4 - w_4^2) } y.
  \EEQ
In this case,  $m$ and $c$ are finite and real, provided $w_4$ is not too large. Further
\BEQ \label{mfg12}
m^2+\half c^2+\frac{\alb}{ y^2}=1 ,
\qquad \Gam_0=\frac{(1+y)^2}{2y^2}\Big[m^2y^2-\alc \Big] .
\EEQ
 Combining them yields
\BEQ
(2\alp+3\alc)m^2+\alc c^2=2\alc  +(2\alp+\alc)\frac{2\Gam_0}{(1+y)^2} .
\EEQ
The expressions (\ref{m2c2v2ha}) for $m^2$ and $c^2$ lead to a cubic equation for $y$,
\BEQ
y=\frac{4(u_4v_4-w_4^2) [ \alc + (2\alp+\alc) {\Gam_0}/ {(1+y)^2}] }{(2\alp+3\alc)(2  v_4  -  w_4)+\alc  (u_4- 2w_4) } ,
\EEQ
which is solved explicitly at the transition line $\Gam_0=0$. 
The remaining relation
\BEQ
\frac{ 2  v_4  -  w_4 }{ 2 (u_4 v_4 - w_4^2 )}=\frac{\alc}{y^3}  + \frac{2\Gam_0}{y(1+y)^2},
\EEQ
is a condition between the various parameters. 
At $\Gam_0=0$ it codes the onset of the spin glass phase.
For small $\Gam_0$ it can be iterated around $\Gam_0=0$. 

Let us investigate the $(\alp,\alc)$ regime where the $T=0$ glass phase exists at given parameters $u_4$, $v_4$ and not-too-large $w_4$, keeping $v_2=\half$. 
For $\alc=0$, eq. (\ref{ergodonT0}) obviously shows that $\Gam_0>0$, this is the glass phase of ref. B.

Eq. (\ref{m2c2v2ha}) in linear in $y$, viz.   $m^2+\half c^2\equiv A y$.
Eq. (\ref{mfg12}) takes the form
\BEQ \label{Ay3}
\hspace{-5mm}
Ay^3-y^2 +\alb,
\quad  A=\frac{ u_4+4  v_4  - 4 w_4 }{ 4 (u_4 v_4 - w_4^2 )} 
,\quad \alb=\alp+\half\alc . 
\EEQ
This equation has a real positive solution for   $\alb<\alb_c= 4/27A^2$, $y<y_c=2/3A$, where $\d y/\d \alb$ is finite.
  The Hessian of $f_0$ as function of $(\qe,m,c)$ has one negative eigenvalue and two positive ones, 
  as it should since $\qe$ is a (spin) glass parameter (recall that $\sum q_\ab^2/n=(n-1)\qe^2\to-\qe^2$ when $q_{\alp\neq\bet}=\qe$.)
   At $\alb_c$, one of the positive eigenvalues goes through zero.
Hence the $T=0$ glass phase exist for $\alb<\alb_c$.
In the standard case $\alc=\alp$, where each pattern possibly has correlation,  it  thus exist for $\alp\le 8/81A^2$.

Next to the case (\ref{m2c2v2ha}) of finite $m$ and $c$, there  exists the stable case $m^2=y/u_4$ and $c=0$.
Eq. (\ref {Ay3})  then holds with  $A\to A'=1/u_4$. Since $A'<A$, it allows a larger region for $2\alp+\alc$,
as $c^2$ does not ``eat way'' from $m^2$ in (\ref{mfg12}a).

\subsection{The spin glass at $T=0$}

For small $T$ one sets 
\BEQ
q_d=1-r_dT,\qquad x(q)=T \xi(q), \qquad \Ie=T\bar\Ie  , \qquad 
\It=T\bar\It .
\EEQ
For $T\to 0$ it brings the scaled functions
\BEQ &&
\bar \Ie(q)=r_d+\int_q^1\d \qb\,\xi(\qb), \qquad \qad \bar\Je(q)=1-\bar \Ie(q) ,
\nn\\&&
 \bar \It(q)=2r_d+\int_q^1\d \qb\,2\qb \xi(\qb), \quad \bar\Jt(q)=\frac{1}{v_2}-\bar \It(q) .
\EEQ
Eq. (\ref{fphysRSB}) reduces to 
\BEQ \label{fphysRSBT0}
 f = && - \frac{u_2}{2}m^2-\frac{u_4}{4}m^4 -\frac{v_2}{2}c^2-\frac{v_4}{4}c^4 -\frac{w_4}{2}m^2c^2 
 +\frac{m^2}{2\bar \Ie(q_0)}+\frac{c^2}{2\bar I_2(q_0)}
\\&&  \hspace{-9mm} \nn
  \mmin \frac{\alp }{2}\Big\{ \frac{q_0}{\bar\Je(q_0)}    \pplus \int_{q_0}^{ 1 }\frac{\d q}{\bar \Je(q)  }  \Big \} 
  \mmin \frac{\alc }{2}\Big\{ \frac{q_0^2}{\bar \Jt(q_0)}   \pplus  \int_{q_0}^{ 1 }\frac{\d q\,2q}{\bar \Jt(q)  }  \Big \}  
  \mmin \frac{1}{2}\Big\{  \frac{q_0}{\bar \Ie(q_0)}   \pplus \int_{q_0}^{ 1 }\frac{\d q}{ \bar \Ie(q) }  \Big\} .
\EEQ
The conditions (\ref{Zpq0}) and (\ref{Zsq0}) become
\BEQ \label{Zpq0T0}
 \frac{2c^2}{\bar \It^2(q_0)}  +  \frac{\alp }{ \bar \Jesq(q_0) } +\frac{6\alc q_0^2}{\bar \Jtsq(q_0) }     =\frac{1 }{\bar \Ie^2(q_0  ) }  
,\qquad 
\frac{m^2}{\bar \Ie^2(q_0)}  =\frac{4\alc q_0^3}{ \bar \Jtsq(q_0) }  .
\EEQ 
Finally, eq. (\ref{xq=}) becomes
\BEQ \label{xqT0=}
\xi(q)=q\times \frac{ 6 \alc  \bar\Ie^3(q) \bar \Je^3(q) \bar \Jt(q) } {  \bar \Je^3(q )\bar \Jt^3(q) + \alp \bar \Ie^3(q) \bar \Jt^3(q) + 8 \alc q^3 \bar \Ie^3(q)  \bar \Je^3(q)}  .
\EEQ

It is seen that, also for the spin glass, the $T\to 0$ limit is well defined.
 The scaled Parisi function $\xi(q)$ remains broad. To avoid $T\log T$ corrections to the free energy, the spherical spins have to be quantized 
 \cite {nieuwenhuizen1995exactly,nieuwenhuizen1995quantum}.

\renewcommand{\thesection}{\arabic{section}.}
\section{Discussion and outlook}
\setcounter{equation}{0}
\renewcommand{\theequation}{\thesection.\arabic{equation}}
\renewcommand{\thesection}{\arabic{section}}

In this work,  a Hopfield model is considered for patterns with internal structure described by random Gaussian variables $\xi^\mu_{ij}$
that describe a correlation between spin $i$ and spin $j$ in pattern $\mu$ for $1\le\mu\le p_c$, where $\alc=p_c/N$ is the loading capacity for 
patterns with structure, out of the $p=\alp N$ total number of patterns. The natural case is $\alc=\alp$, but it is instructive to consider general $\alc$.
The employed spherical limit for the spins makes the problem solvable analytically. 
Correlations due to structures in patterns can be incorporated in many neural networks, such as those  with Ising spins or Potts spins \cite{bolle1992thermodynamic}.

While the spherical Hopfield model has a glass phase but no spin glass phase, the correlations in the structures
induce a spin glass state with continuous (full)  replica symmetry breaking.
It is the state entered when coming from high temperature, and likely occurs also for Ising spins. 
At zero temperature, a glass phase exists for $2\alp+\alc$ below a certain threshold, while a spin glass phase exist above this threshold. 
Phases with metastable or stable patterns and/or correlations exist, partly separated by first order phase transitions.
While various properties of the phases have been elucidated,  a construction of the full phase diagram is left as a task for future.

{\color{black} Another} task for future is to study the dynamics of the network. One can extend the Langevin dynamics of the spins treated in Boll\'e et al. \cite{bolle2003spherical}
to include Langevin forces for the correlation parameters  $\xi^\mu_{ij}$.
 
 {\color{black} One may also consider higher order structures in the patterns, leading to correlations between triplets, quartets, quintets, etc, of spins.
 The quartet case is of special interest, given the shape of cloths, floors, city plans, etc.}

Another future direction  is the following.
If, in a spin glass, the disorder couplings are not quenched but evolve slowly in time in an equilibrium state at a temperature $T_A$,  
the replica number becomes finite, $n=T/T_A$ \cite{penney1993coupled}. 
This situation was worked out for the $p$-spin spin glass model \cite{allahverdyan2000model}. Such models are related to machine learning,
to which correlation in patterns can be added.
In the present case, the patterns $\xi_i^\mu$ and structures $\xi^\mu_{ij}$ can have very different long timescales and be related to an equilibrium 
at temperatures different  from each other and different from $T$, so that even two finite replica numbers can appear. 
An application of this approach is a double-replica theory for the evolution of genotype-phenotype interrelationship \cite{pham2023double}.

\renewcommand{\thesection}{\arabic{section}.}
\section{Appendix A: Replica theory for spherical models}
\setcounter{equation}{0}
\renewcommand{\theequation}{A.\arabic{equation}}
\renewcommand{\thesection}{\arabic{section}}

Crisanti and Sommers in 1992 have presented the eigenvalues of the Parisi overlap matrix $q_{\alp\bet}$ with $K$ steps of RSB \cite{crisanti1992spherical}.
Here we calculate the quantity  $\log\det q$ and related expressions as function of $n$ and take the llmit $n\to0$. 

To define the $n\to 0$ replica limit, 
Parisi considers the class of integers $n$ which decomposes as the product $n=m_1 m_2\cdots m_K$ with integers $n\ge m_i\ge 1$ such that 
$m_i/m_{i+1}\ge 1$ is also an  integer.. With the $K$ steps of replica symmetry breaking, $m_0=n$ and  $m_{K+1}=1$, 
the overlaps $q_\ab$ take the values
\BEQ 
q_\ab=q_i \qquad \text{if}\quad  
I(\alp/m_i) = I(\bet/m_i) \quad \text{and}\quad I(\alp/m_{i+1} ) \neq   I(\bet/m_{i+1}). \quad
\EEQ  
where $I(x) = \lceil x \rceil$ denotes the smallest integer greater than or equal to $x$.
The inverse Parisi function $x(q)$ is (often $q(x)$ is called Parisi function)
\BEQ
x(q)
&=&n+\sum_{i=0}^K (m_{i+1}-m_i)\theta(q-q_i)  \nn\\
&=&n\theta(q_0-q)+m_{1}\theta(q-q_0) +\sum_{i=1}^K (m_{i+1}-m_i)\theta(q-q_i) , 
\EEQ
with the property $x(q)=1$ for $q_K\le q \le 1$. Special case are $K=0$ (replica symmetry) where $x(q)=n+(1-n)\theta(q-q_0)$ and $K=1$ (1RSB) where
$x(q)=n+(m_1-n)\theta(q-q_0)+(1-m_1)\theta(q-\qe )$. 

The eigenvalues of $q_{\alpha\beta}$ and their degeneracies are 
\BEQ 
&& \Ie(0) =\int_0^1\d q\,x(q) \, \, =nq_0+\Ie(q_0) ,\qquad \qquad \qquad  {\rm deg}=1,          \nn \\ 
&& \Ie(q_0) =\int_{q_0}^1\d q\,x(q) =1-q_K+ \int_{q_0}^{q_K}\d q\,x(q)  ,  \qquad  \! {\rm deg} =\frac{n}{m_1}-1 , 
 \\
 &&  \Ie(q_i)  =\int_{q_i}^1\d q\,x(q) =1-q_K+ \int_{q_i}^{q_K}\d q\,x(q)  , \qquad  {\rm deg}=\frac{n}{m_{i+1}}-\frac{n}{m_i} , 
  \nn
  \\
&&   \Ie(q_K)=1-q_K ,  \qquad \hspace{43mm} {\rm deg}=n -\frac{n}{m_K} .
\EEQ
The total number of eigenvalues  is $n/m_{K+1}=n$, as it should.
It holds that 
\BEQ \label{Iqi}
 \Ie(q_i) &=& 1-m_{i+1}q_i-(1-m_K) q_K  -  \sum_{j=i+1}^{K-1} (m_{j+1}-m_j) q_j ,  
  \nn\\ 
&=& 1-q_K +  \sum_{j=i+1}^K m_j(q_j -q_{j-1}) , \qquad \quad (i=0,\cdots ,K-1),
\EEQ
implying the recursion 
\BEQ\label{Iqrec}
\Ie(q_{i-1}) =\Ie(q_i)+m_{i}(q_{i}-q_{i-1})  , \qquad
 \qquad (i=1,\cdots , K) .
\EEQ
The non-degenerate eigenvalue
\BEQ
I_0=1+(n-m_1)q_0+ (m_K-1)q_K+ \sum_{j=1}^{K-1} (m_j-m_{j+1}) q_j 
\EEQ
relates to the eigenvector with equal entries, $e_\alp^{(1)}=1/\sqrt{n}$,  and exhibits the blocks of $q_{\alp\bet}$ of size $(m_{j+1}-m_j)\times (m_{j+1}-m_j)$ 
with value $q_j$, while the terms $m_{K+1}=1$  relate to the diagonal ``block'' of size $m_{K+1}\times m_{K+1}$, expressing the diagonal elements $q_{\alp\alp}=1$. 
The other eigenvectors are perpendicular to this one, so $\sum_\alp e^{(j)}_\alp=0$ for $j=2,\cdots,n$. Hence only  $e_\alp^{(1)}$ contributes to 
\BEQ \label{sumqinvab} \hspace{-3mm}
\sum_{\alp\beta}(q^{-1})_{\alp\bet} \iss \frac{n}{\Ie(0)} \iss \frac{n}{nq_0 \pplus \Ie(q_0)}
, \hspace{3mm}
\sum_{\alp\beta}(Q^{-1})_{\alp\bet}=\frac{n}{\It(0)} \iss \frac{n}{nq_0^2\pplus I_2(q_0)}.
\EEQ

This leads to the exact result at finite $n$,
\BEQ 
\hspace{-3mm}
\log\det q = \log \frac{\Ie(q_0)+nq_0}{\Ie(q_0)} +n\log \Ie(q_K) + n\sum_{i=1}^{K} \frac{\log [\Ie(q_{i-1})/ \Ie(q_i)]}{m_{i}}.   
\EEQ
The $n\to0$ limit leads to the free energy with $K$ steps of replica symmetry breaking,
\BEQ \label{detqK}
\hspace{-3mm}
\frac{\log\det q}{n} 
=\frac{q_0}{\Ie(q_0)} +\log \Ie(q_K) + \sum_{i=1}^{K} \frac{\log [\Ie(q_{i-1})/ \Ie(q_i)]}{m_{i}}.   
\EEQ
where the order of the $m_i$ gets reversed, $n\le m_i\le m_{i+1}\le 1$, with the $m_i$ now being real numbers.

Physical phases with a continuous part and various steps were found in models for 
site disordered magnets \cite{nieuwenhuizen1999theory}.
Solutions with a finite number ($K=1,2,\cdots$) of replica symmetry breaking steps  were not  found  in those models, 
but they were discovered in $s+p$ spherical models \cite{crisanti2007amorphous}.

In the continuum limit $K\to\infty$,  the $q_i$ become dense between $q_0$ and $\qe $, with renamed  $\qe \equiv q_K$.
Setting $m_i\to x(q_i)$, one gets for a smooth function $x(q)$
\BEQ 
\frac{\log\det q}{n} 
=\frac{q_0}{\Ie(q_0)} +\log (1-\qp) + \int_\qn^\qp\frac{ \d q}{\Ie(q)} , 
\EEQ
which involves, from (\ref{Iqi}),  the continuum function
\BEQ \label{Iqis}
\hspace{-4mm}
\Ie(q) \iss 1 \mmin \qe  \pplus \qe x(q_d)  \mmin q x(q)  \mmin \int _{x(q)}^{x(\qe )}\d x \, q(x)
\iss 1 \mmin \qe  \pplus \int _{q}^{\qe }\d \qb\, x(\qb)  .
\EEQ

 Subtraction of a diagonal matrix $T\veen$ with $T=1/\bet$ immediately leads to 
\BEQ   \hspace{-5mm}
\frac{\log\det (\veen-\bet q)}{n} 
=\log [1-\bet(1-\qp)] -\frac{q_0}{ T-\Ie(q_0)} - \int_\qn^\qp\frac{ \d q} { T-  \Ie(q)}  . 
\EEQ

For the matrix $Q$ with elements $Q_{\alp\bet}=(q_{\alp\bet})^2$ one employs the same block sizes $m_i$
and replaces $q_i\to Q_i=q_i^2$.  With $\d Q=\d q\,2q$ it thus follows that 
\BEQ  \hspace{-5mm}
\frac{\log\det(1-\bet_2 Q)}{n} =
\log [1-\bet_2(1-\qe ^2)]  \mmin  \frac{q_0^2}{ T_2 \mmin \It(q_0)}  \mmin \int_\qn^\qp\frac{ \d q\, \,2 q}{T_2 \mmin \It(q) } ,
\EEQ
for $\beta_2=1/T_2=v_2/T$. Analogous to (\ref{Iqis}) one has
\BEQ
I_2(q) =  1- \qe ^2 +\int_q^{\qe }\d \qb\, 2 \qb x(\qb). 
\EEQ
Finally, eq. (\ref{sumqinvab}) implies for $n\to 0$
\BEQ
\frac{1}{n}\sum_{\alp\beta}(q^{-1})_{\alp\bet}=\frac{1}{\Ie(q_0)}, \qquad
\frac{1}{n}\sum_{\alp\beta}(Q^{-1})_{\alp\bet}=\frac{1}{I_2(q_0)} .
\EEQ

\hspace{3mm} 

\renewcommand{\thesection}{\arabic{section}.}
\section{Appendix B: Fluctuations in the glass state}
\setcounter{equation}{0}
\renewcommand{\theequation}{B.\arabic{equation}}
\renewcommand{\thesection}{\arabic{section}}

\newcommand{\wa}{\sqrt{\alpha}}

\newcommand{\bm}{\bar m}
\newcommand{\mb}{\bar m}
\newcommand{\cb}{\bar c}

To find the eigenvalues of the stability matrix in the replica symmetric glass phase with magnetization and correlation order parameters, 
we connect to Appendix A of Crisanti and Leuzzi  for the Blume-Emery-Griffiths-Capel, that also has two magnetization-type order parameters \cite{crisanti2004thermodynamic}.

The starting point is the free energy functional (\ref{fnfin}),
\BEQ  \label{fnfin2}
 f_n &=&  \sum_\alp \Big[
 - \frac{u_2}{2} m_\alp^2-\frac{ u_4}{4} m_\alp^4
 -\frac{v_2}{2}  c_\alp^2  - \frac{ v_4}{4}  c_\alp^4  -\frac{w_4}{2}m_\alp^2c_\alp^2\Big] \nn \\ && 
 +\frac{T}{2} \sum_{\alp\bet} \Big [ m_\alp   (q^{-1})_{\alp\bet}  m_\bet  +  c_\alp  (Q^{-1})_{\alp\bet} c_\bet\Big] 
\\&&  +\frac{\alp T}{2} \log \det (\veen-\beta_1 q) + \frac{\alc T}{2} \log \det (\veen-\bet_2 Q)  - \frac{T}{2} \log \det q . \nn
\EEQ
 For the $\alp$ and $\alc$ terms we need the combinations
\BEQ
q_T=q_d+T_1,\qquad Q_T=q_d^2+T_2,\quad T_1=\frac{T}{u_2},\quad T_2=\frac{T}{v_2},
\EEQ
while the $m_\alp m_\bet$ and $c_\alp c_\bet$ terms involve $q_0$ and $Q_0$, respectively.
The inverses of the unperturbed replica symmetric matrices read
\BEQ 
&& \hspace{-9mm}  
{\bf g}_T  \iss \frac{\veen}{(1 \mmin \qT)\veen \pplus \qe\vE } \iss s_T (\veen-r_T \vE) , \qad
  s_T =\frac{1}{1-\qT}  , \quad  
 r _T \iss \frac{\qe}{1\mmin \qT+n\qe} , \qad
 \\ &&  \hspace{-9mm} 
\vG_T =\frac{\veen}{(1 \mmin Q_T )\veen \pplus \qe^2\vE }
=S_T (\veen \mmin R_T  \vE) , \quad 
S _T \iss\frac{1}{1-Q_T}  , \qad  R_T  \iss \frac{\qe^2}{1\mmin Q_T +n\qe^2} . \nn
\EEQ
Let us define for any scalar S, vector $V$ and matrix $M$ 
\BEQ
(S)=S,\quad 
 [V]=\sum_\alp V_\alp, \quad 
(M)=\sum_\alp M_{\alp\alp}, \quad [M]=\sum_\ab M_\ab .
\EEQ

\newcommand{\zmm}{6mm}

Up to second order, using $(\qb)=(\Qb)=0$ and $[\qb_2]=(\qb^2)$, $(\veen)=(\vE)=n$
there occur their quadratic and bilinear fluctuation terms 
\BEQ   \label{F2=}
\hspace{-\zmm} 
\delta^2 f_n=&&
-\half \{ u_2+3u_4m^2+w_4c^2\}(\bm\bm)
-\half \{ v_2+3v_4m^2+w_4m^2\}(\cb\cb)-2w_4mc(\bm\cb)
\nn\\&&  \hspace{-\zmm} 
+ \frac{T}{2}\frac{s_0^2}{2}\{  (\qb^2)- 2r_0[\qb^2]+r_0^2[\qb]^2\} 
-\frac{\alp T}{2}\frac{s_T^2}{2}\{  (\qb^2)- 2r_T[\qb^2]+r_T^2[\qb]^2\} 
\nn\\&&  \hspace{-\zmm} 
-\alc T \qe^2S_T^2 \{ (\qb^2) -  2R_T[\qb^2] + R_T^2[\qb]^2\} -\frac{\alc T}{2} S_T R_T (\qb^2) 
 \nn\\ &&  \hspace{-\zmm} 
+ \frac{T}{2} s_0 \{  (\mb^2) -r_0[\mb]^2 \}
+\frac{T}{2} m^2s_0^3(1-nr_0)^2 \{ [\qb^2]-r_0[\qb]^2\}  
 \\ &&  \hspace{-\zmm} 
+\frac{T}{2} S_0\{(\bar c^2)-R_0[\bar c]^2\}
+\frac{T}{2} c^2S_0^2(1-nR_0)^2\{4\qe^2S_0\{[\qb^2]-R_0[\qb]^2\}  -(\qb^2)  \}  
 \nn\\ &&  \hspace{-\zmm} 
-s_0^2(1-nr_0)m\{[\qb\mb]-r_0[\qb][\mb] \} 
-2\qe S_0^2(1-nR_0)c\{[\qb\bar c]-R_0[\qb][\bar c] \} . \nn
\EEQ
These terms can be gathered as
\BEQ && 
2\delta^2f_n=\bar A_2 (\bq^2)+\bar A_1[\bq^2]+\bar  A_0[\bq]^2
+\bar  B_1(\cb\cb)+\bar  B_0[\cb]^2
\nn \\ &&
+ \bar  C_1(\bm\bm)+\bar  C_0[\bm]^2 +\bar  D_1([\bq\cb]+[\cb\bq])+\bar  D_0([\bq][\cb]+[\cb][\bq])
\\ &&+\bar  E_1([\bq\bm]+[\bm\bq])+\bar  E_0([\bq][\bm]+[\bm][\bq])+F_1[(\bm\bar c)+(\bar c\bm)]. \nn
\EEQ
For $n\to 0$ the coefficients read
\BEQ
&&
\bar A_2=\frac{T}{2(1\mmin \qe)^2}
 \mmin \frac{c^2T}{(1\mmin \qe^2)^2}
 \mmin \frac{\alp T} {2 (T\mmin 1\pplus\qe)^2 } 
\mmin \frac{3\alc \qe^2 T }{(T_2\mmin 1\pplus\qe^2)^2 }  \,  ,
 \nn\\ &&
\bar A_1=  \frac{\alp  q_d T}{(1 \mmin q_d \mmin T)^3} 
 \pplus  \frac{4 \alc  q_d^4 T}{(1  \mmin q_d^2 \mmin T_2)^3} 
\mmin \frac{c^2 q_d^2 T}{(1 \mmin q_d^2)^4} 
 \mmin   \frac{m^2 q_d T}{(1 \mmin q_d)^4} \mmin \frac{q_d T}{(1 \mmin q_d)^3} \, ,
 \nn\\ &&
\bar A_0=  \frac{q_d^2 T}{2 (1 \mmin q_d)^4} 
\mmin \frac{\alp  q_d^2 T}{2 (1 \mmin q_d \mmin T)^4}  
  \mmin \frac{2 \alc  q_d^6 T}{(1 \mmin q_d^2 \mmin T_2)^4}  \, ,
 \nn\\ &&
\bar B_1 =   \frac{T}{1 \mmin q_d^2} \mmin 3 m^2 v_4 \mmin m^2 w_4  \mmin v_2 ,  \quad \bar B_0=  \mmin \frac{q_d^2 T}{(1 \mmin q_d^2)^2} \, ,
 \nn\\ &&
\bar C_1=  \frac{T}{1 \mmin q_d} \mmin 3 m^2 u_4   \mmin c^2 w_4  \mmin u_2 , \quad \,\, \bar C_0= \mmin \frac{ q_d T}{(1 \mmin q_d)^2} \, ,
 \nn\\ &&
\bar D_1=\frac{2 c q_d T}{\left(1 \mmin q_d^2\right)^2} ,\quad 
\bar D_0=  \mmin \frac{2 c q_d^3 T}{ (1 \mmin q_d^2 )^3} ,
\nn \\ && 
\bar E_1= \frac{m T}{(1 \mmin q_d)^2}  ,\quad \,
\bar E_0= \mmin \frac{ m q_d T}{(1 \mmin q_d)^3}  \, ,
\nn\\ && 
\bar F_1= \mmin 2 c  m w_4, \quad \, \, \bar F_0 = 0   \, . 
\EEQ
The Crisanti-Leuzzi parameters are
\BEQ &&
A_2=\bar A_2+2\bar A_1+\bar A_0,\quad A_1=\bar A_1+\bar A_0,\qad A_0=\bar A_0 
\nn \\ && 
B_1=\bar B_1+\bar B_0 ,\quad 
B_0=\bar B_0 ,\quad 
C_1=\bar C_1+\bar C_0 ,\quad 
C_0=\bar C_0 ,\quad 
\\ && 
D_1=\bar D_1+\bar D_0 ,\qad 
D_0=\bar D_0 ,\quad 
E_1=\bar E_1+\bar E_0 ,\qad  
E_0=\bar E_0 ,\quad
F_1 \, =\bar F_1 .
\EEQ

Now let us denote
\BEQ
A=A_2-4A_1+3A_0=\bar A_2-2\bar A_1 .
\EEQ
The eigenvalues of the longitudinal fluctuations (non-degenrate) and the anomalous fluctuations (degenerate) coincide for $n\to 0$, 
\BEQ \label{Lammin}
\Lambda_\pm=A+\bar B_1 \pm \sqrt{(A-\bar B_1)^2-8\bar D_1^2} . 
\EEQ
 The potentially dangerous one is $\Lambda_-$. 
 For $T\to 0$, $\Lambda_+\sim +1/T^2$ but, using $q_d=1-T/(y+1)$, $\Lambda_-$ appears to remain finite,
\BEQ \label{Lammin0}
\Lambda_-=y+1- 2 v_2 -2 m^2 (3 v_4 + w_4) .
\EEQ
The replicon eigenvalue is
 \BEQ \label{Gam=}
\Lambda_R=\Gam =
\frac{1}{2(1\mmin \qe)^2}
 \mmin \frac{c^2}{(1\mmin \qe^2)^2}
 \mmin \frac{\alp } {2 (T\mmin 1\pplus\qe)^2 } 
\mmin \frac{3\alc \qe^2  }{(T_2\mmin 1\pplus\qe^2)^2 }   .
\EEQ
The high temperature phase $q_d=m=c=0$  is unstable below $T_\sg=1+\sqrt{\alp}$.
When $q_d\neq 0$,  the mean field equation (\refl{mfBolleQ}) leads to the simpler form
\BEQ
\Gam = \frac{1}{2\qe}\Big[ \frac{m^2}{(1 - \qe)^2}  -  \frac{4\alc \qe^3 }{(T_2-1 + \qe^2 )^2} \Big].
    \EEQ
For $T\to0$ this becomes also $\sim 1/T^2$, 
\BEQ
\Gam = \frac{(y+1)^2}{2T^2}\Big[ m^2  -  \frac{4\alc v_2^2 }{(y+1-2v_2)^2} \Big].
\EEQ
The condition $\Gam \downarrow0$ sets  a smooth transition from a glass phase to a spin glass phase.
For $\alc=0$, there is no region with $\Gam<0$, hence no spin glass  \cite{bolle2003spherical}.
For $v_2=\half $ there results
\BEQ
\Gam = \frac{(y+1)^2}{2T^2y^2}\Big( m^2y^2  -  \alc \Big).
\EEQ

Data Availability Statement: No Data associated in the manuscript.

\section{References}



\end{document}